\documentclass[
			preprint,
			12pt,authoryear,
			3p,
			times,
			]{elsarticle}

\usepackage{amssymb}
\usepackage{amsmath}

\usepackage{hyperref}
\usepackage{eurosym}
\usepackage{xcolor}

\journal{Journal of Plant Physiology}

\begin{document}

\begin{frontmatter}



\title{A multi-physics approach to probing plant responses:\\ From calcium signaling to thigmonastic motion} 

\author[DTU]{Sabrina Gennis}
\author[DTU]{Matthew D. Biviano}
\author[DTU]{Kristoffer P. Lyngbirk}
\author[JIC,Zhejiang]{Hannah R. Thomas}
\author[WSU]{Viktoriya Vasina}
\author[JIC]{Christine Faulkner}
\author[WSU]{Michael Knoblauch}
\author[DTU]{Kaare H. Jensen}
\ead{khjensen@fysik.dtu.dk}

\affiliation[DTU]{organization={Department of Physics, Technical University of Denmark},
             city={2800 Kongens Lyngby},
             country={Denmark}}
\affiliation[JIC]{organization={Cell and Developmental Biology, John Innes Centre},
             city={Norwich NR4 7UH},
             country={United Kingdom}}
\affiliation[Zhejiang]{organization={Current Address: Department of Horticulture, Zijingang Campus, Zhejiang University},
             city={866 Yuhangtang Road, Hangzhou, 310058},
             country={China}}
\affiliation[WSU]{organization={School of Biological Sciences, Washington State University},
             city={Pullman WA 99164},
             country={USA}}

\begin{abstract}
Plants respond to biotic and abiotic stresses through complex and dynamic mechanisms that integrate physical, chemical, and biological cues. Here, we present a multi-physics platform designed to systematically investigate these responses across scales. The platform combines a six-axis micromanipulator with interchangeable probes to deliver precise mechanical, electrostatic, optical, and chemical stimuli. Using this system, we explore calcium signaling in \emph{Arabidopsis thaliana}, thigmonastic motion in \emph{Mimosa pudica}, and chemical exchange via microinjection in \emph{Rosmarinus officinalis L.} and \emph{Ocimum basilicum}.
Our findings highlight stimulus-specific and spatially dependent responses: mechanical and electrostatic stimuli elicit distinct calcium signaling patterns, while repeated electrostatic stimulation exhibited evidence of response fatigue. Thigmonastic responses in \emph{Mimosa pudica} depend on the location of perturbation, highlighting the intricate bi-directional calcium signaling. Microinjection experiments successfully demonstrate targeted chemical perturbations in glandular trichomes, opening avenues for biochemical studies.
This open-source platform provides a versatile tool for dissecting plant stress responses, bridging the gap between fundamental research and applied technologies in agriculture and bioengineering. By enabling precise, scalable, and reproducible studies of plant-environment interactions, this work offers new insights into the mechanisms underlying plant resilience and adaptability.
\end{abstract}



\begin{keyword}
Micromanipulation \sep Plant Signaling \sep Calcium \sep Thigmonastic motion \sep Microinjection \sep Plant Biomechanics \sep Plant Biophysics




\end{keyword}

\end{frontmatter}


\section{Introduction}\label{sec:Introduction}
Plants are immobile, and much of their existence is shaped by a struggle to persist in harsh environments while resisting the transfer of assimilated carbon to other terrestrial lifeforms \citep{farmer2014leaf}. Hidden behind the shroud of seeming pacificity and immobility, plants' response to biotic and abiotic stresses are therefore, perhaps unsurprisingly, extremely complicated and multi-layered \citep{erb2019molecular}.
Illustrative biotic examples include, e.g., herbivore \citep{acevedo2015cues, Abbas2017VolatileTerpenoids} and pathogen \citep{saur2021recognition} perception, and lesser explored topics, such as, gall production \citep{stone2003adaptive}, resistance to oviposition \citep{williams1981insects}, and mutualistic interactions with ants \citep{erb2019molecular}. Abiotic factors are equally diverse, from light to drought, salinity and temperature \citep{zhang2022abiotic}.

Each of the aforementioned challenges is unique and must, whenever possible, be studied in isolation to minimize potentially confounding effects. This approach has been applied across physiology, ecology, stress, pathology, and bioengineering (see, e.g., \citep{Farmer2020ElectricalSignals, Lee2022WoundInduced}).  However, while the downstream effects are distinctive, the basic mechanical principles of the triggers share some common features \citep{pedley2000blood}. In other words, the majority of the stimulants can be broken down into a limited set of basic physical interactions. These include, for instance, mechanical, chemical, optical, and electrostatic triggers. Refocusing on plants' response to these basic stimuli could potentially shed new light on the fundamental underlying processes.

However, a clear technical knowledge gap presents itself: no experimental system exists for studying multi-physical biotic and abiotic stress triggers in plants. Here, we report such a system: a six-axis micromanipulator equipped with interchangeable attachments. The probes allow us to explore mechanical, chemical, optical, and electrostatic triggers. We applied this to induce chemical signaling in \emph{Arabidopsis thaliana}, motion in \emph{Mimosa pudica}, and targeted deposition of fluorescent molecules in the glandular trichomes of \emph{Rosmarinus officinalis L.} and \emph{Ocimum basilicum}. The setup allows us to compare and contrast the effects of different probes. Additionally, we can explore the effects of timing in repeated triggering through a simple programming interface.

The paper is structured as follows. Initially, we present the physical probes and the micromanipulation system, followed by a discussion on software integration. In the Applications section, we first examine how touch, light, and electrical stimulation of  \emph{Arabidopsis} elicits various responses in calcium signaling. Next, we focus on \emph{Mimosa} and highlight the significance of stimulus location. Lastly, we demonstrate the injection of fluorescent molecules into glandular trichomes of basil and rosemary. We provide detailed information about the experimental system, and the software is available on GitHub, along with files for 3D printing: \url{http://www.autoplant.org/AUTOPLANT/tree/main/Multi-Physics}.

\begin{figure*}
\includegraphics[]{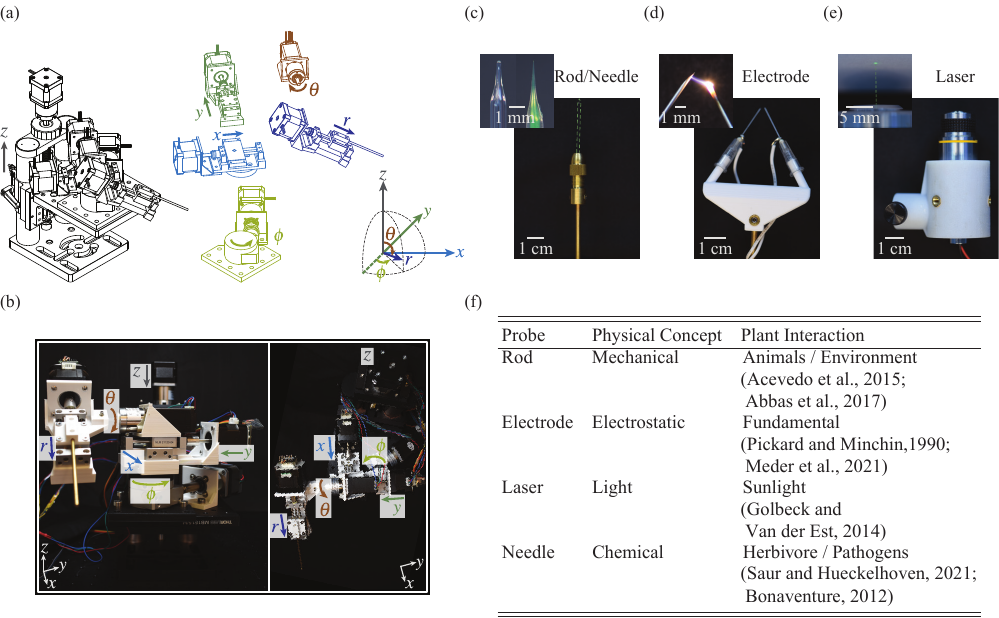}
\caption{\label{fig:Nectar} Multi-physics approach to probing plant responses. (a) Assembled and exploded-view drawings of the micromanipulator. Motion along the six axes (translational: $x$, $y$, $z$, and $r$; rotational: $\theta$ and $\phi$) is controlled separately. The probes are mounted to the $r$ or $\theta$ axes. (b) Photograph of the experimental setup. (c) Mechanical and chemical stimuli are generated using the rod and needle probes. (d) The electrode probe triggers electrical perturbations. (e) A focused laser beam mimics optical interactions. (f) An overview of the physical concept, probe, and corresponding plant interactions.}
\end{figure*}

\section{Experimental setup}
\subsection{Probes}
\label{sec:probes}

To systematically explore triggers of biotic and abiotic stress in plants, we designed four different probes, each representative of a unique physical interaction: mechanical, electrostatic, optical, and chemical (Fig. \ref{fig:Nectar}c-e). For an overview of the probes, the underlying physical concept they explore, and examples of plant interactions, see Fig. \ref{fig:Nectar}f. Briefly, we used a glass rod (tip diameter 0.15 mm) to mimic mechanical contact. A pair of electrodes (tip diameter 0.1 mm, spacing 0.7-1.3 mm, voltage difference $\sim$ 6.75 kV) were used to explore the effects of an electrical field. A diode laser (520 nm, 5 V or 12.5 V, spot size 0.5-2 mm, electrical power input 6 mW or 11 mW, maximum irradiance 8 kW/m$^2$) provided a localized optical perturbation. Lastly, a glass needle (tip diameter 0.02 mm) filled with a fluorescent dye was used to study chemical interaction. We note that additional probes simulating, for instance, temperature or wind could be developed; however, we will not do that here. Details of the probe design and fabrication are provided in the Materials and Methods section at the end of the manuscript.

\subsection{Micromanipulator}
\label{sec:micromanipulator}

Plants' responses to stimuli depend not only on the type of perturbation but also on the location and timing of the impulse. Therefore, we attached the previously mentioned probes to an electromechanical device capable of simultaneous micrometer positioning and probe control. Our system is an extension of well-established micromanipulator technology. However, while commercially available micromanipulators are precise and come with varying degrees of geometric freedom \citep{Zhang2019, Adam2021Micromanipulation}, they do not typically integrate directly with software for imaging and triggering required for our applications. It is worth noting that this issue is not exclusive to plants, as micromanipulation is also crucial in numerous other fields, including neuroscience \citep{Ryan2020NeuroscienceEquipment}, in vitro fertilization \citep{Wijegunawardana2021MEMSIVF}, and microinjection \citep{Pan2024RoboticMicroinjection,Abdelrahman2021Microinjection}, among others.

Therefore, we designed and built an open-source, integrated micromanipulator, triggering, and imaging system using a mixture of 3D-printed and widely available parts (Fig. \ref{fig:Nectar}a,b). Probes are attached to a generic sample holder and are positioned with micrometer resolution along six geometric axes. Probe alignment, activation, and data collection (imaging) are integrated into a single software package. The system can reach a resolution of about 4 $\mu$m in the individual axes. However, care has to be taken when switching directions, and the backlash of up to 80 $\mu$m has to be corrected for. To our best knowledge, there are only a few micromanipulators in the scientific community that are open-source, affordable, offer a high resolution, and a travel range on the macro scale \citep{Hietanen2018OpenSourceProbe}. In addition, our system is motorized and equipped with open-source software. Details of the setup are provided in the Materials and Methods section.

\section{Results}
Stress responses in plants are remarkably sophisticated. However, our hypothesis is that some stress responses can be classified based on the applied basic physical stimuli, the triggering of which may provide useful model systems. To demonstrate the feasibility of this idea, this section explores responses and compares different stimuli and their timings to garner insights into their effects.

\begin{figure*}[htb]
\includegraphics[]{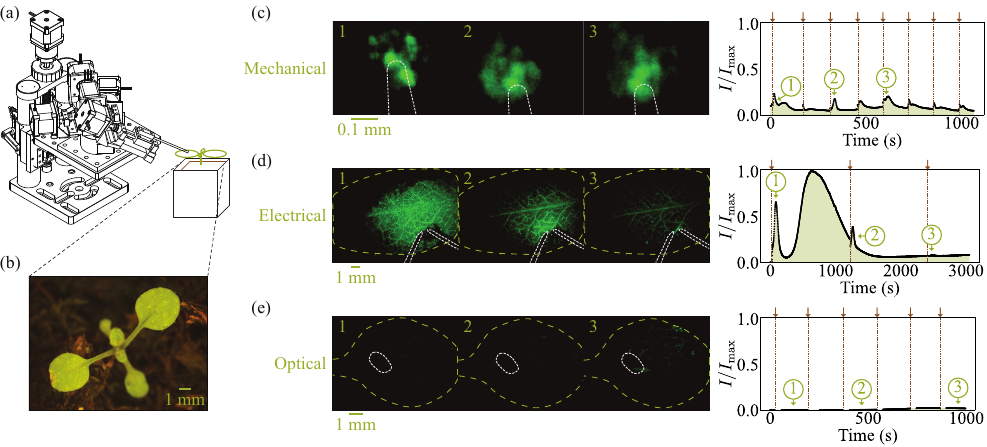}
\caption{\label{fig:CalciumResponse} The qualitative and quantitative characteristics of calcium signaling in \textit{Arabidopsis thaliana} mutants expressing \textit{GCaMP3} in response to mechanical, electrical, and optical stimuli. (a-b) Experimental setup. (c-e) Fluorescence imaging was employed to investigate the spatiotemporal dynamics of signals. The fluorescent intensity $I$, averaged over the field-of-view, was quantified as a function of time $t$. Vertical arrows indicate stimulus timing. (c) A mechanical stimulus (a rod contacting the leaf surface) triggered a calcium signal at the contact point. Still images captured 15 seconds post-touch demonstrate the typical response, which is local and relatively short-lived. A similar reaction was observed when the touch was repeated. (d) An electrical disturbance caused by an electrode above the leaf produced a significantly stronger calcium signal. Images taken 60 seconds post-application illustrate the typical response, which is relatively prolonged and affects nearly the entire leaf. Importantly, the response to repeated stimuli diminished. The cause of this apparent fatigue remains uncertain, but it does not seem to be due to irreversible tissue damage (Fig. \ref{fig:CalciumBeforeAfter}). (e) Lastly, the reaction to the optical stimulus from a focused laser beam is similar to that of the control experiment (no stimulus, Fig. \ref{fig:CalciumControl}). Data in (c-e) are available in the supplemental materials (Video A1-A3).}
\end{figure*}

\subsection{Calcium Signalling}
\label{sec:calcium}

Calcium release is one of the initial responses of plants to a diverse range of stimuli
\citep{Rasul2017, bellandi2022glutamate, Fichman2021IntegrationSignals}. We conduct experiments to quantify the strength and timing of this stress response following three basic physical perturbations: mechanical, electrical, and optical stress. The \textit{Arabidopsis thaliana} transgenic line \textit{35S::GCaMP3} provides a convenient platform for this experiment; a fluorescent signal reveals how the cellular calcium concentration varies in space and time. Several previous studies have explored this process, with \textit{GCaMP} being first developed by \citet{Nakai2001CaProbe} and improved upon by \citet{Tian2009ImagingNeuralActivity}. The transgenic line has been used in studies with \textit{Arabidopsis thaliana} to untangle the effects of wounding on the whole plant by investigating systemic responses \citep{toyota2018glutamate}. Members of our team have previously used this line to elucidate the transport mechanism of amino acid chemical messengers \citep{bellandi2022glutamate} and to determine how pavement cells distinguish touch from letting go \citep{Howell2023PavementCells}. However, the link between the type of stimulus the plant is exposed to, the timing of the perturbation, and the plant's response remains unclear. Here, using different (repeated) stimuli, we show how the plant's response differs, while a deeper analysis of the stimuli, including different probe sizes and strengths, should be discussed in future studies.

Touching the leaf using the rod probe (Fig. \ref{fig:CalciumResponse}a,b) resulted in a localized signal that propagated radially outwards from the point of contact (Fig. \ref{fig:CalciumResponse}c). A ring-like structure spread from the source at a speed of approximately $0.01$ mm/s to a range of about $0.25$ mm, consistent with previous observations \citep{bellandi2022glutamate, Howell2023PavementCells}. The signal strength (i.e., integrated optical intensity) peaked at approximately $15$ s after the first contact. Repeating the stimulus at intervals of $120$ s elicited a similar response. Although the signal magnitude did vary between contacts, we did not find systematic evidence of desensitization (fatigue \citep{pickard1990phloem}). Notably, the rod's tip was $\sim 0.15$ mm wide; hence, only a few epidermal cells directly sensed the mechanical contact.

Contrast this with the response to an electrical stimulus generated by the electrode probe (Fig. \ref{fig:CalciumResponse}d). Here, the initial signal spread faster (up to $\sim 1.3$ mm/s at first, then the speed decreased as the signal spreads) and traveled further (range several cm), eventually reaching the petiole. From here, the signal propagated to distal leaves, and a second wave could be observed due to the returning signal creating an echo in the initially perturbed region. One reason why the electrode probe elicited a stronger response than was achieved by mechanical contact could be that it is larger, thus initially triggering many epidermal cells at once (recall that the electrode spacing is $\sim 1 $ mm). The response to subsequent stimuli, however, was weakened, an effect that has previously been observed for stimuli such as cold shock \citep{Minchin1983RateCooling}, vibration \citep{pickard1990phloem}, wind \citep{Knight1992WindInduced}, or the gravitational field \citep{Plieth2002ReorientationSeedlings}. Even after waiting ten times longer ($1200$ s) than between mechanical stimuli ($120$ s), the magnitude of the second response was only $40 \%$ of the primary response, while a third attempt (after $2400$ s) yielded a response reduced to $\sim 1\%$ (Fig. \ref{fig:CalciumResponse}d). Other studies have investigated plant defense responses to flame wounding \citep{Vian1999ChloroplastMRNA, Hlavackova2006ElectricalChemicalSignals}, which could present a possible explanation for the observed fatigue due to cell death. Although we have neither observed any permanent fluorescently inactive regions nor any tissue damages at this magnification (Fig. \ref{fig:CalciumBeforeAfter}a,b), we can not rule out cell death as a reason for the decrease in signal. However, when repeating the experiment on a plant 2.5 hours after the initial stimulation, the same pattern as before was observed (Fig. \ref{fig:CalciumBeforeAfter}c,d). This suggests recovery occurs between experiments and that the reduction in response over shorter timeframes is a result of fatigue rather than irreversible tissue damage or cell death, though the cause of the reduced sensitivity remains unclear and should be explored in future studies.

\begin{figure*}[ht!]
\includegraphics[]{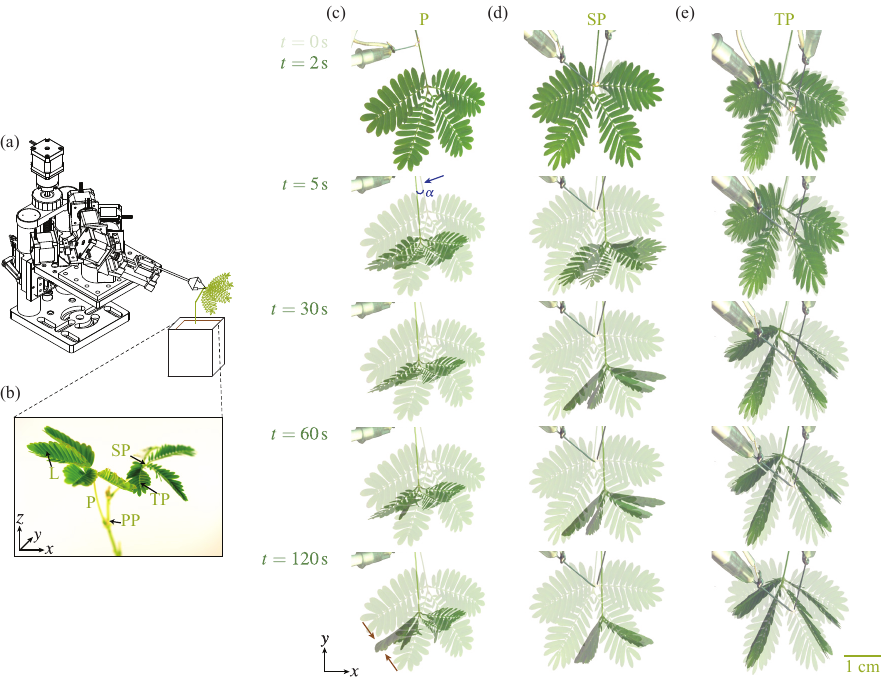}
\caption{\label{fig:Mimosa} Thigmonastic movement in \emph{Mimosa pudica} and the impact of stimulus placement. (a-b) Experimental setup: The electrode probe is placed near the petiole (P), secondary pulvinus (SP), and tertiary pulvinus (TP). (c-e) Sequential images of Mimosa pudica in top view during and after stimulation. (c) When the petiole is disturbed, it pivots within seconds (angle $\alpha$), and only after 120 seconds do the leaves droop (arrows). (d) Stimulation of the secondary pulvinus results in both effects occurring almost simultaneously. (e) When the tertiary pulvinus is disturbed, drooping leaves are observed, yet the petiole retains its turgor. Data in (c-e) are available in the supplemental materials (Video A4-A6).}
\end{figure*}

A third set of experiments explored the effects of a weak optical stimulus using the laser probe, but this did not result in a clear calcium response (Fig. \ref{fig:CalciumResponse}e). The signal was consistent with background measurements without stimuli (Fig. \ref{fig:CalciumControl}), and the weak laser did not cause any visible tissue damage. Increasing the laser power from $5$ to $12.5$ V, however, left burn marks, a process which has been widely studied previously \citep{Vian1999ChloroplastMRNA, Hlavackova2006ElectricalChemicalSignals}, and which we will therefore not discuss in detail here.

\subsection{Thigmonastic motion}
\label{sec:thigmo}

After investigating how various physical stimuli applied to the same location can elicit different responses, we turned our attention to how the spatial positioning of disturbances influences the reaction. We utilized \emph{Mimosa pudica} as our model system, a plant known for its compound leaves that quickly droop when stimulated, a well-known instance of thigmonastic motion \citep{Weintraub1952LeafMovements}. The rapid leaf movements are mediated by changes in calcium concentration \citep{Hagihara2022CalciumMediated} (c.f., Section \ref{sec:calcium}).

\textit{Mimosa pudica} does not just react to touch but has also been observed to respond to electrical stimulation if electrodes are embedded in parts of the plant \citep{Volkov2010MimosaPudica}. However, it has not been investigated how \textit{Mimosa pudica} reacts to an external electric field. Here, we used the electrode probe, which can be positioned freely, to explore this. 

An electrical arc was applied in close proximity to either the petiole, the secondary pulvinus, or a tertiary pulvinus of \textit{Mimosa pudica} (Fig. \ref{fig:Mimosa}b). (Neither the probe nor the actual electrical arc made contact with the plant.) 
When applied to the petiole, the initial reaction was turgor loss in the primary pulvinus, which caused the petiole to pivot (Fig. \ref{fig:Mimosa}c, angle $\alpha$, at 5 s). The motion stopped ($\Delta \alpha \approx 45-60^\circ$) after approximately $\sim 10$ s. Subsequently, the leaflets closed, a process completed $\sim 120$ s after. The change in angle on the primary pulvinus seems consistent with observations made for a dark/light cycle of the plant \citep{Song2014PulvinusBending}.
If the secondary pulvinus was exposed to an electrical stimulus, the petiole also lost turgor, and the leaflets folded. However, the timing was different: When perturbing the secondary pulvinus, leaflet closure ($\sim 2$ s) began before turgor loss of the petiole was initiated. Notably, the steady-state ($\Delta \alpha \approx 35-60^\circ$, leaflets closed) was reached more than three times faster ($\sim 30$ s). This is not taking into account the subsequent turgor loss of the secondary pulvinus visible at $\sim 120$ s, which was only observed in some instances. Finally, we applied the electrode probe at a tertiary pulvinus. The leaves again folded up, but turgor in the petiole was retained. The stationary state ($\Delta \alpha =0^\circ$, leaflets closed) again took $\sim 30$ s to materialize.

These simple observations illustrate how the spatial positioning of disturbances influence the response, which appears consistent with a bi-directionality signal emanating from the perturbed region \citep{Hagihara2022CalciumMediated}. The contact-free electrical triggering system introduced here could provide a convenient tool for future exploration of signal propagation in thigmonastic motion.
  
The electrode probe induced no visible damage on the plant. Therefore, we assume that all responses in the thigmonastic motion resulted not from a burn, but from the electrostatic interaction. Using a high-power laser probe (12.5 V) on a leaflet or tertiary pulvini resulted in a similar dynamic as observed with the electrode probe in Fig. \ref{fig:Mimosa}c. However, in addition, a burn mark was visible after illumination, which has been studied previously by, e.g., \citet{Malone1994WoundInduced} and \citet{Volkov2013MorphingStructures}.

\begin{figure*}[htb!]
\includegraphics[]{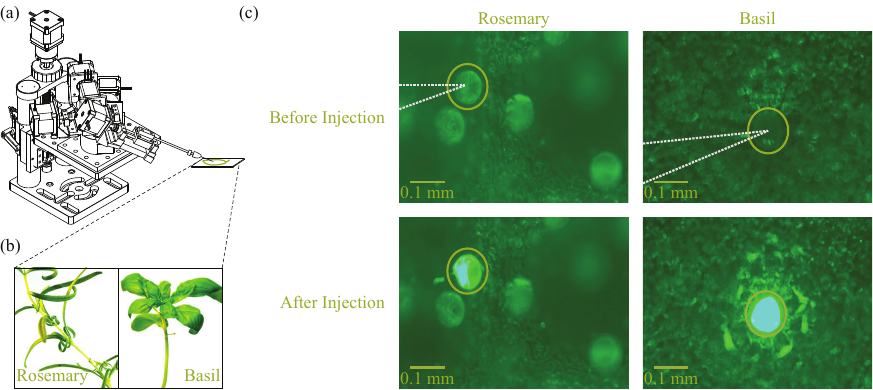}
\caption{\label{fig:Microinjection}Chemical stimulus. (a-b) Experimental setup. A narrow glass capillary tube with a tip diameter of 0.02 mm was secured to the micromanipulator for microinjecting fluorescein dye into the glandular trichomes of rosemary and basil. (c) Fluorescent images show the glands before and after the injection. The dashed line indicates the needle. Circles highlight the glandular trichomes.}
\end{figure*}

\subsection{Microinjection}
\label{sec:microinjection}

Having established the feasibility of exploring mechanical, electrical, and optical stimuli, we now turn our attention to a final example: chemical exchange. This mode of perturbation is inspired by aphids (and sharpshooters), insects that use their piercing mouthparts to feed on plant sap \citep{Bonaventure2012PerceptionInsectFeeding}. Saliva injected into plants by aphids acts as a calcium scavenger to suppress wound response reactions \citep{will2007molecular} (c.f., Sec. \ref{sec:calcium} and \ref{sec:thigmo}).

Our system is not yet able to fully mimic aphids, which are able to target individual cells in vascular tissues hidden deep within the plant. However, to demonstrate the feasibility of injection into individual cells (Fig. \ref{fig:Microinjection}a,b), we targeted glandular trichomes, organs located on the leaf surface. We used a narrow glass capillary tube (tip diameter 0.02 mm) to microinject a fluorescent dye (fluorescein) into the glandular trichomes of rosemary and basil. We observed the infiltration of the dye, which makes the targeted trichomes fluoresce brighter under a microscope compared to before injection (Fig. \ref{fig:Microinjection}c). In basil, the surrounding epidermal cells also show increased fluorescence, possibly due to the trichomes being more embedded in the epidermis, making the surrounding cells reflect the light.

\section{Discussion and conclusion}
Plants are exposed to varying biotic and abiotic stressors \citep{Abbas2017VolatileTerpenoids, zhang2022abiotic} which are triggering complex processes \citep{erb2019molecular}. Some of these plant-stressor interactions have been well studied \citep{Rasul2017, bellandi2022glutamate, Fichman2021IntegrationSignals}. However, a technical gap persists in investigating multi-physical biotic and abiotic triggers in plants.

Motivated to fill a part of this gap in experimental methodologies, we introduced a novel multi-physics platform that enables precise investigation of plant responses to biotic and abiotic stress. The development of four distinct probes—mechanical, electrostatic, optical, and chemical—allowed for the systematic exploration of fundamental physical interactions that underlie complex stress responses in plants (Sec. \ref{sec:probes} and Fig. \ref{fig:Nectar}c-f). By integrating these probes with a versatile, open-source six-axis micromanipulator (Sec. \ref{sec:micromanipulator} and Fig. \ref{fig:Nectar}a,b), this platform provides a powerful tool for probing plant physiology at scales ranging from the cellular to the macroscopic scale.

The calcium signaling experiments in \emph{Arabidopsis thaliana} (Sec. \ref{sec:calcium}) highlighted nuanced differences in plant responses to mechanical and electrostatic perturbations, with evidence of response fatigue observed during repeated electrical stimulation (Fig. \ref{fig:CalciumResponse}d). This finding indicates that plants exhibit stress-specific fatigue mechanisms, providing insights into how plants manage repeated or prolonged exposure to stimuli. In contrast, the optical stimulus did not elicit a calcium response under the tested conditions (Fig. \ref{fig:CalciumResponse}e), potentially highlighting limitations in plant perception of specific wavelengths or intensities of light. The multi-physics system allows for future studies to systemically test probes by varying the lasers light intensity and wavelength, the rods touch strength and the voltage difference of the electrode.

The investigation of thigmonastic motion in \emph{Mimosa pudica} (Sec. \ref{sec:thigmo}) underscored the spatial dependence of stimulus responses. Variations in turgor loss and leaflet closure patterns, depending on the stimulus location, emphasized the intricate signaling pathways involved in plant motion (Fig. \ref{fig:Mimosa}c). The ability to apply localized stimuli without physical contact offers exciting opportunities for exploring signal propagation dynamics in plant tissues.

Finally, microinjection of a fluorescent dye into glandular trichomes of rosemary and basil (Sec. \ref{sec:microinjection}) showcased the platform's potential for chemical perturbations and application of components to individual structures at a microscale. This capability opens avenues for studying biochemical exchanges in plant tissues with high precision.

Collectively, these findings demonstrate the versatility and utility of the platform, which bridges scales and disciplines. By dissecting plant responses to fundamental physical interactions, this study lays the groundwork for a deeper understanding of how plants perceive and respond to their environments.

The platform's open-source design facilitates replication and customization, and together with its low cost of approximately \euro3500, broad accessibility for the research community can be achieved. Future work could expand the system’s capabilities, such as integrating additional probes for temperature, wind, or acoustic stimulation. Moreover, combining multiple stimuli in tandem could reveal interactions between different signaling pathways, providing a holistic view of plant stress responses.

\section*{Acknowledgments}
This work was supported by the Independent Research Fund Denmark [grant number 1127-00219B]; the Biotechnology and Biological Science Research Council [grant numbers BB/X010996/1, BB/X007685/1]; the European Research Council [grant number 725459 - “INTERCELLAR"]; and the NSF-IOS [2318280].

\section*{Declaration of generative AI and AI-assisted technologies in the writing process.}
During the preparation of this work the author(s) used Grammarly in order to check spelling and grammar. After using this tool/service, the author(s) reviewed and edited the content as needed and take(s) full responsibility for the content of the publication.

\nocite{Golbeck2014BiophysicsPhotosynthesis} 
\nocite{Meder2021UltraconformableElectrodes} 

\bibliographystyle{elsarticle-harv} 
\bibliography{Bib}

\begin{thebibliography}{42}
\expandafter\ifx\csname natexlab\endcsname\relax\def\natexlab#1{#1}\fi
\providecommand{\url}[1]{\texttt{#1}}
\providecommand{\href}[2]{#2}
\providecommand{\path}[1]{#1}
\providecommand{\DOIprefix}{doi:}
\providecommand{\ArXivprefix}{arXiv:}
\providecommand{\URLprefix}{URL: }
\providecommand{\Pubmedprefix}{pmid:}
\providecommand{\doi}[1]{\href{http://dx.doi.org/#1}{\path{#1}}}
\providecommand{\Pubmed}[1]{\href{pmid:#1}{\path{#1}}}
\providecommand{\bibinfo}[2]{#2}
\ifx\xfnm\relax \def\xfnm[#1]{\unskip,\space#1}\fi
\bibitem[{Abbas et~al.(2017)Abbas, Ke, Yu et~al.}]{Abbas2017VolatileTerpenoids}
\bibinfo{author}{Abbas, F.}, \bibinfo{author}{Ke, Y.}, \bibinfo{author}{Yu,
  R.}, et~al., \bibinfo{year}{2017}.
\newblock \bibinfo{title}{Volatile terpenoids: multiple functions,
  biosynthesis, modulation and manipulation by genetic engineering}.
\newblock \bibinfo{journal}{Planta} \bibinfo{volume}{246},
  \bibinfo{pages}{803--816}.
\newblock \URLprefix \url{https://doi.org/10.1007/s00425-017-2749-x},
  \DOIprefix\doi{10.1007/s00425-017-2749-x}.
\bibitem[{Abdelrahman et~al.(2021)Abdelrahman, Hasan and
  Da'as}]{Abdelrahman2021Microinjection}
\bibinfo{author}{Abdelrahman, D.}, \bibinfo{author}{Hasan, W.},
  \bibinfo{author}{Da'as, S.I.}, \bibinfo{year}{2021}.
\newblock \bibinfo{title}{Microinjection quality control in zebrafish model for
  genetic manipulations}.
\newblock \bibinfo{journal}{MethodsX} \bibinfo{volume}{8},
  \bibinfo{pages}{101418}.
\newblock \DOIprefix\doi{10.1016/j.mex.2021.101418}.
\bibitem[{Acevedo et~al.(2015)Acevedo, Rivera-Vega, Chung, Ray and
  Felton}]{acevedo2015cues}
\bibinfo{author}{Acevedo, F.E.}, \bibinfo{author}{Rivera-Vega, L.J.},
  \bibinfo{author}{Chung, S.H.}, \bibinfo{author}{Ray, S.},
  \bibinfo{author}{Felton, G.W.}, \bibinfo{year}{2015}.
\newblock \bibinfo{title}{Cues from chewing insects—the intersection of
  damps, hamps, mamps and effectors}.
\newblock \bibinfo{journal}{Current Opinion in Plant Biology}
  \bibinfo{volume}{26}, \bibinfo{pages}{80--86}.
\bibitem[{Adam et~al.(2021)Adam, Chidambaram, Reddy, Ramani and
  Cappelleri}]{Adam2021Micromanipulation}
\bibinfo{author}{Adam, G.}, \bibinfo{author}{Chidambaram, S.},
  \bibinfo{author}{Reddy, S.}, \bibinfo{author}{Ramani, K.},
  \bibinfo{author}{Cappelleri, D.}, \bibinfo{year}{2021}.
\newblock \bibinfo{title}{Towards a comprehensive and robust micromanipulation
  system with force-sensing and vr capabilities}.
\newblock \bibinfo{journal}{Micromachines} \bibinfo{volume}{12},
  \bibinfo{pages}{784}.
\newblock \DOIprefix\doi{10.3390/mi12070784}.
\bibitem[{Bae et~al.(2021)Bae, Paludan, Knoblauch and
  Jensen}]{Bae2021NeuralNetworks}
\bibinfo{author}{Bae, H.}, \bibinfo{author}{Paludan, M.},
  \bibinfo{author}{Knoblauch, J.}, \bibinfo{author}{Jensen, K.H.},
  \bibinfo{year}{2021}.
\newblock \bibinfo{title}{Neural networks and robotic microneedles enable
  autonomous extraction of plant metabolites}.
\newblock \bibinfo{journal}{Plant Physiology} \bibinfo{volume}{186},
  \bibinfo{pages}{1435--1441}.
\newblock \URLprefix \url{https://doi.org/10.1093/plphys/kiab178},
  \DOIprefix\doi{10.1093/plphys/kiab178}.
\bibitem[{Bellandi et~al.(2022)Bellandi, Papp, Breakspear, Joyce, Johnston,
  de~Keijzer, Raven, Ohtsu, Vincent, Miller, Sanders, Hogenhout, Morris and
  Faulkner}]{bellandi2022glutamate}
\bibinfo{author}{Bellandi, A.}, \bibinfo{author}{Papp, D.},
  \bibinfo{author}{Breakspear, A.}, \bibinfo{author}{Joyce, J.},
  \bibinfo{author}{Johnston, M.G.}, \bibinfo{author}{de~Keijzer, J.},
  \bibinfo{author}{Raven, E.C.}, \bibinfo{author}{Ohtsu, M.},
  \bibinfo{author}{Vincent, T.R.}, \bibinfo{author}{Miller, A.J.},
  \bibinfo{author}{Sanders, D.}, \bibinfo{author}{Hogenhout, S.A.},
  \bibinfo{author}{Morris, R.J.}, \bibinfo{author}{Faulkner, C.},
  \bibinfo{year}{2022}.
\newblock \bibinfo{title}{Diffusion and bulk flow of amino acids mediate
  calcium waves in plants}.
\newblock \bibinfo{journal}{Science Advances} \bibinfo{volume}{8},
  \bibinfo{pages}{eabo6693}.
\newblock \DOIprefix\doi{10.1126/sciadv.abo6693}.
\bibitem[{Bonaventure(2012)}]{Bonaventure2012PerceptionInsectFeeding}
\bibinfo{author}{Bonaventure, G.}, \bibinfo{year}{2012}.
\newblock \bibinfo{title}{Perception of insect feeding by plants}.
\newblock \bibinfo{journal}{Plant Biology} \bibinfo{volume}{14},
  \bibinfo{pages}{872--880}.
\newblock \URLprefix \url{https://doi.org/10.1111/j.1438-8677.2012.00650.x},
  \DOIprefix\doi{10.1111/j.1438-8677.2012.00650.x}. \bibinfo{note}{epub 2012
  Sep 7}.
\bibitem[{Erb and Reymond(2019)}]{erb2019molecular}
\bibinfo{author}{Erb, M.}, \bibinfo{author}{Reymond, P.}, \bibinfo{year}{2019}.
\newblock \bibinfo{title}{Molecular interactions between plants and insect
  herbivores}.
\newblock \bibinfo{journal}{Annual review of plant biology}
  \bibinfo{volume}{70}, \bibinfo{pages}{527--557}.
\bibitem[{Farmer(2014)}]{farmer2014leaf}
\bibinfo{author}{Farmer, E.E.}, \bibinfo{year}{2014}.
\newblock \bibinfo{title}{Leaf defence}.
\newblock \bibinfo{publisher}{OUP Oxford}.
\bibitem[{Farmer et~al.(2020)Farmer, Gao, Lenzoni, Wolfender and
  Wu}]{Farmer2020ElectricalSignals}
\bibinfo{author}{Farmer, E.E.}, \bibinfo{author}{Gao, Y.Q.},
  \bibinfo{author}{Lenzoni, G.}, \bibinfo{author}{Wolfender, J.L.},
  \bibinfo{author}{Wu, Q.}, \bibinfo{year}{2020}.
\newblock \bibinfo{title}{Wound- and mechanostimulated electrical signals
  control hormone responses}.
\newblock \bibinfo{journal}{New Phytologist} \bibinfo{volume}{227},
  \bibinfo{pages}{1037--1050}.
\newblock \DOIprefix\doi{10.1111/nph.16646}.
\bibitem[{Fichman and Mittler(2021)}]{Fichman2021IntegrationSignals}
\bibinfo{author}{Fichman, Y.}, \bibinfo{author}{Mittler, R.},
  \bibinfo{year}{2021}.
\newblock \bibinfo{title}{Integration of electric, calcium, reactive oxygen
  species and hydraulic signals during rapid systemic signaling in plants}.
\newblock \bibinfo{journal}{The Plant Journal} \bibinfo{volume}{107},
  \bibinfo{pages}{7--20}.
\newblock \URLprefix \url{https://doi.org/10.1111/tpj.15360},
  \DOIprefix\doi{10.1111/tpj.15360}.
\bibitem[{Golbeck and Van~der Est(2014)}]{Golbeck2014BiophysicsPhotosynthesis}
\bibinfo{editor}{Golbeck, J.}, \bibinfo{editor}{Van~der Est, A.} (Eds.),
  \bibinfo{year}{2014}.
\newblock \bibinfo{title}{The Biophysics of Photosynthesis}. volume
  \bibinfo{volume}{465}.
\newblock \bibinfo{publisher}{Springer}, \bibinfo{address}{New York}.
\bibitem[{Hagihara et~al.(2022)Hagihara, Mano, Miura
  et~al.}]{Hagihara2022CalciumMediated}
\bibinfo{author}{Hagihara, T.}, \bibinfo{author}{Mano, H.},
  \bibinfo{author}{Miura, T.}, et~al., \bibinfo{year}{2022}.
\newblock \bibinfo{title}{Calcium-mediated rapid movements defend against
  herbivorous insects in mimosa pudica}.
\newblock \bibinfo{journal}{Nature Communications} \bibinfo{volume}{13},
  \bibinfo{pages}{6412}.
\newblock \DOIprefix\doi{10.1038/s41467-022-34106-x}.
\bibitem[{Hietanen et~al.(2018)Hietanen, Heikkinen, Savin and
  Pearce}]{Hietanen2018OpenSourceProbe}
\bibinfo{author}{Hietanen, I.}, \bibinfo{author}{Heikkinen, I.T.},
  \bibinfo{author}{Savin, H.}, \bibinfo{author}{Pearce, J.M.},
  \bibinfo{year}{2018}.
\newblock \bibinfo{title}{Approaches to open source 3-d printable probe
  positioners and micromanipulators for probe stations}.
\newblock \bibinfo{journal}{HardwareX} \bibinfo{volume}{4}.
\newblock \URLprefix \url{https://doi.org/10.1016/j.ohx.2018.e00042},
  \DOIprefix\doi{10.1016/j.ohx.2018.e00042}. \bibinfo{note}{accessed 2025 Jan
  16}.
\bibitem[{Hlav{\'a}{\v{c}}kov{\'a} et~al.(2006)Hlav{\'a}{\v{c}}kov{\'a},
  Krch{\v{n}}{\'a}k, Nau{\v{s}}
  et~al.}]{Hlavackova2006ElectricalChemicalSignals}
\bibinfo{author}{Hlav{\'a}{\v{c}}kov{\'a}, V.},
  \bibinfo{author}{Krch{\v{n}}{\'a}k, P.}, \bibinfo{author}{Nau{\v{s}}, J.},
  et~al., \bibinfo{year}{2006}.
\newblock \bibinfo{title}{Electrical and chemical signals involved in
  short-term systemic photosynthetic responses of tobacco plants to local
  burning}.
\newblock \bibinfo{journal}{Planta} \bibinfo{volume}{225},
  \bibinfo{pages}{235--244}.
\newblock \DOIprefix\doi{10.1007/s00425-006-0325-x}.
\bibitem[{Howell et~al.(2023)Howell, Völkner, McGreevy
  et~al.}]{Howell2023PavementCells}
\bibinfo{author}{Howell, A.H.}, \bibinfo{author}{Völkner, C.},
  \bibinfo{author}{McGreevy, P.}, et~al., \bibinfo{year}{2023}.
\newblock \bibinfo{title}{Pavement cells distinguish touch from letting go}.
\newblock \bibinfo{journal}{Nature Plants} \bibinfo{volume}{9},
  \bibinfo{pages}{877--882}.
\newblock \DOIprefix\doi{10.1038/s41477-023-01418-9}.
\bibitem[{Knight et~al.(1992)Knight, Smith and
  Trewavas}]{Knight1992WindInduced}
\bibinfo{author}{Knight, M.R.}, \bibinfo{author}{Smith, S.M.},
  \bibinfo{author}{Trewavas, A.J.}, \bibinfo{year}{1992}.
\newblock \bibinfo{title}{Wind-induced plant motion immediately increases
  cytosolic calcium}.
\newblock \bibinfo{journal}{Proceedings of the National Academy of Sciences of
  the United States of America} \bibinfo{volume}{89},
  \bibinfo{pages}{4967--4971}.
\newblock \URLprefix \url{https://doi.org/10.1073/pnas.89.11.4967},
  \DOIprefix\doi{10.1073/pnas.89.11.4967}.
\bibitem[{Lee and Seo(2022)}]{Lee2022WoundInduced}
\bibinfo{author}{Lee, K.}, \bibinfo{author}{Seo, P.J.}, \bibinfo{year}{2022}.
\newblock \bibinfo{title}{Wound-induced systemic responses and their
  coordination by electrical signals}.
\newblock \bibinfo{journal}{Frontiers in Plant Science} \bibinfo{volume}{13}.
\newblock \DOIprefix\doi{10.3389/fpls.2022.880680}.
\bibitem[{Malone(1994)}]{Malone1994WoundInduced}
\bibinfo{author}{Malone, M.}, \bibinfo{year}{1994}.
\newblock \bibinfo{title}{Wound‐induced hydraulic signals and stimulus
  transmission in mimosa pudica l.}
\newblock \bibinfo{journal}{New Phytologist} \bibinfo{volume}{128},
  \bibinfo{pages}{49--56}.
\bibitem[{Meder et~al.(2021)Meder, Saar, Taccola, Filippeschi, Mattoli and
  Mazzolai}]{Meder2021UltraconformableElectrodes}
\bibinfo{author}{Meder, F.}, \bibinfo{author}{Saar, S.},
  \bibinfo{author}{Taccola, S.}, \bibinfo{author}{Filippeschi, C.},
  \bibinfo{author}{Mattoli, V.}, \bibinfo{author}{Mazzolai, B.},
  \bibinfo{year}{2021}.
\newblock \bibinfo{title}{Ultraconformable, self-adhering surface electrodes
  for measuring electrical signals in plants}.
\newblock \bibinfo{journal}{Advanced Materials Technologies}
  \bibinfo{volume}{6}, \bibinfo{pages}{2001182}.
\newblock \DOIprefix\doi{10.1002/admt.202001182}.
\bibitem[{Minchin and Thorpe(1983)}]{Minchin1983RateCooling}
\bibinfo{author}{Minchin, P.E.H.}, \bibinfo{author}{Thorpe, M.R.},
  \bibinfo{year}{1983}.
\newblock \bibinfo{title}{A rate of cooling response in phloem translocation}.
\newblock \bibinfo{journal}{Journal of Experimental Botany}
  \bibinfo{volume}{34}, \bibinfo{pages}{529--536}.
\bibitem[{Nakai et~al.(2001)Nakai, Ohkura and Imoto}]{Nakai2001CaProbe}
\bibinfo{author}{Nakai, J.}, \bibinfo{author}{Ohkura, M.},
  \bibinfo{author}{Imoto, K.}, \bibinfo{year}{2001}.
\newblock \bibinfo{title}{A high signal-to-noise ca2+ probe composed of a
  single green fluorescent protein}.
\newblock \bibinfo{journal}{Nature Biotechnology} \bibinfo{volume}{19},
  \bibinfo{pages}{137--141}.
\newblock \DOIprefix\doi{10.1038/84397}.
\bibitem[{Pan et~al.(2024)Pan, Zoberman, Zhang
  et~al.}]{Pan2024RoboticMicroinjection}
\bibinfo{author}{Pan, P.}, \bibinfo{author}{Zoberman, M.},
  \bibinfo{author}{Zhang, P.}, et~al., \bibinfo{year}{2024}.
\newblock \bibinfo{title}{Robotic microinjection enables large-scale transgenic
  studies of caenorhabditis elegans}.
\newblock \bibinfo{journal}{Nature Communications} \bibinfo{volume}{15},
  \bibinfo{pages}{8848}.
\newblock \DOIprefix\doi{10.1038/s41467-024-53108-5}.
\bibitem[{Pedley(2000)}]{pedley2000blood}
\bibinfo{author}{Pedley, T.}, \bibinfo{year}{2000}.
\newblock \bibinfo{title}{Blood flow in arteries and veins}.
\newblock \bibinfo{journal}{Perspectives in fluid dynamics: a collective
  introduction to current research} , \bibinfo{pages}{105--158}.
\bibitem[{Pickard and Minchin(1990)}]{pickard1990phloem}
\bibinfo{author}{Pickard, W.F.}, \bibinfo{author}{Minchin, P.E.H.},
  \bibinfo{year}{1990}.
\newblock \bibinfo{title}{{The Transient Inhibition of Phloem Translocation in
  Phaseolus vulgaris by Abrupt Temperature Drops, Vibration, and Electric
  Shock}}.
\newblock \bibinfo{journal}{Journal of Experimental Botany}
  \bibinfo{volume}{41}, \bibinfo{pages}{1361--1369}.
\newblock \DOIprefix\doi{10.1093/jxb/41.11.1361}.
\bibitem[{Plieth and Trewavas(2002)}]{Plieth2002ReorientationSeedlings}
\bibinfo{author}{Plieth, C.}, \bibinfo{author}{Trewavas, A.J.},
  \bibinfo{year}{2002}.
\newblock \bibinfo{title}{Reorientation of seedlings in the earth's
  gravitational field induces cytosolic calcium transients}.
\newblock \bibinfo{journal}{Plant Physiology} \bibinfo{volume}{129},
  \bibinfo{pages}{786--796}.
\newblock \URLprefix \url{https://doi.org/10.1104/pp.011007},
  \DOIprefix\doi{10.1104/pp.011007}.
\bibitem[{Rasul et~al.(2017)Rasul, Nadeem, Siddique, Atif, Ali, Umer, Rashid,
  Afzal, Abid and Azeem}]{Rasul2017}
\bibinfo{author}{Rasul, I.}, \bibinfo{author}{Nadeem, H.},
  \bibinfo{author}{Siddique, M.H.}, \bibinfo{author}{Atif, R.M.},
  \bibinfo{author}{Ali, M.A.}, \bibinfo{author}{Umer, A.},
  \bibinfo{author}{Rashid, F.}, \bibinfo{author}{Afzal, M.},
  \bibinfo{author}{Abid, M.}, \bibinfo{author}{Azeem, F.},
  \bibinfo{year}{2017}.
\newblock \bibinfo{title}{Plants sensory-response mechanisms for salinity and
  heat stress}.
\newblock \bibinfo{journal}{Journal of Animal and Plant Sciences}
  \bibinfo{volume}{27}, \bibinfo{pages}{490--502}.
\bibitem[{Ryan et~al.(2020)Ryan, Johnson and
  Deitcher}]{Ryan2020NeuroscienceEquipment}
\bibinfo{author}{Ryan, J.}, \bibinfo{author}{Johnson, B.R.},
  \bibinfo{author}{Deitcher, D.}, \bibinfo{year}{2020}.
\newblock \bibinfo{title}{Building your own neuroscience equipment: A precision
  micromanipulator and an epi-fluorescence microscope for calcium imaging}.
\newblock \bibinfo{journal}{Journal of Undergraduate Neuroscience Education}
  \bibinfo{volume}{19}, \bibinfo{pages}{A134--A140}.
\bibitem[{Saur and H{\"u}ckelhoven(2021)}]{saur2021recognition}
\bibinfo{author}{Saur, I.M.}, \bibinfo{author}{H{\"u}ckelhoven, R.},
  \bibinfo{year}{2021}.
\newblock \bibinfo{title}{Recognition and defence of plant-infecting fungal
  pathogens}.
\newblock \bibinfo{journal}{Journal of Plant Physiology} \bibinfo{volume}{256},
  \bibinfo{pages}{153324}.
\bibitem[{Song et~al.(2014)Song, Yeom and Lee}]{Song2014PulvinusBending}
\bibinfo{author}{Song, K.}, \bibinfo{author}{Yeom, E.}, \bibinfo{author}{Lee,
  S.}, \bibinfo{year}{2014}.
\newblock \bibinfo{title}{Real-time imaging of pulvinus bending in mimosa
  pudica}.
\newblock \bibinfo{journal}{Scientific Reports} \bibinfo{volume}{4},
  \bibinfo{pages}{6466}.
\newblock \URLprefix \url{https://doi.org/10.1038/srep06466},
  \DOIprefix\doi{10.1038/srep06466}.
\bibitem[{Stone and Sch{\"o}nrogge(2003)}]{stone2003adaptive}
\bibinfo{author}{Stone, G.N.}, \bibinfo{author}{Sch{\"o}nrogge, K.},
  \bibinfo{year}{2003}.
\newblock \bibinfo{title}{The adaptive significance of insect gall morphology}.
\newblock \bibinfo{journal}{Trends in Ecology \& Evolution}
  \bibinfo{volume}{18}, \bibinfo{pages}{512--522}.
\bibitem[{Tian et~al.(2009)Tian, Hires, Mao
  et~al.}]{Tian2009ImagingNeuralActivity}
\bibinfo{author}{Tian, L.}, \bibinfo{author}{Hires, S.}, \bibinfo{author}{Mao,
  T.}, et~al., \bibinfo{year}{2009}.
\newblock \bibinfo{title}{Imaging neural activity in worms, flies, and mice
  with improved gcamp calcium indicators}.
\newblock \bibinfo{journal}{Nature Methods} \bibinfo{volume}{6},
  \bibinfo{pages}{875--881}.
\newblock \DOIprefix\doi{10.1038/nmeth.1398}.
\bibitem[{Toyota et~al.(2018)Toyota, Spencer, Sawai-Toyota, Jiaqi, Zhang, Koo,
  Howe and Gilroy}]{toyota2018glutamate}
\bibinfo{author}{Toyota, M.}, \bibinfo{author}{Spencer, D.},
  \bibinfo{author}{Sawai-Toyota, S.}, \bibinfo{author}{Jiaqi, W.},
  \bibinfo{author}{Zhang, T.}, \bibinfo{author}{Koo, A.J.},
  \bibinfo{author}{Howe, G.A.}, \bibinfo{author}{Gilroy, S.},
  \bibinfo{year}{2018}.
\newblock \bibinfo{title}{Glutamate triggers long-distance, calcium-based plant
  defense signaling}.
\newblock \bibinfo{journal}{Science} \bibinfo{volume}{361},
  \bibinfo{pages}{1112--1115}.
\newblock \DOIprefix\doi{10.1126/science.aat7744}.
\bibitem[{Vian et~al.(1999)Vian, Henry-Vian and
  Davies}]{Vian1999ChloroplastMRNA}
\bibinfo{author}{Vian, A.}, \bibinfo{author}{Henry-Vian, C.},
  \bibinfo{author}{Davies, E.}, \bibinfo{year}{1999}.
\newblock \bibinfo{title}{Rapid and systemic accumulation of chloroplast
  mrna-binding protein transcripts after flame stimulus in tomato}.
\newblock \bibinfo{journal}{Plant Physiology} \bibinfo{volume}{121},
  \bibinfo{pages}{517--524}.
\newblock \DOIprefix\doi{10.1104/pp.121.2.517}.
\bibitem[{Volkov et~al.(2010)Volkov, Foster, Ashby, Walker, Johnson and
  Markin}]{Volkov2010MimosaPudica}
\bibinfo{author}{Volkov, A.G.}, \bibinfo{author}{Foster, J.C.},
  \bibinfo{author}{Ashby, T.A.}, \bibinfo{author}{Walker, R.K.},
  \bibinfo{author}{Johnson, J.A.}, \bibinfo{author}{Markin, V.S.},
  \bibinfo{year}{2010}.
\newblock \bibinfo{title}{Mimosa pudica: Electrical and mechanical stimulation
  of plant movements}.
\newblock \bibinfo{journal}{Plant, Cell \& Environment} \bibinfo{volume}{33},
  \bibinfo{pages}{163--173}.
\newblock \DOIprefix\doi{10.1111/j.1365-3040.2009.02066.x}.
\bibitem[{Volkov et~al.(2013)Volkov, O’Neal, Volkova and
  Markin}]{Volkov2013MorphingStructures}
\bibinfo{author}{Volkov, A.G.}, \bibinfo{author}{O’Neal, L.},
  \bibinfo{author}{Volkova, M.I.}, \bibinfo{author}{Markin, V.S.},
  \bibinfo{year}{2013}.
\newblock \bibinfo{title}{Morphing structures and signal transduction in mimosa
  pudica l. induced by localized thermal stress}.
\newblock \bibinfo{journal}{Journal of Plant Physiology} \bibinfo{volume}{170},
  \bibinfo{pages}{1317--1327}.
\newblock \URLprefix \url{https://doi.org/10.1016/j.jplph.2013.05.003},
  \DOIprefix\doi{10.1016/j.jplph.2013.05.003}.
\bibitem[{Weintraub(1952)}]{Weintraub1952LeafMovements}
\bibinfo{author}{Weintraub, M.}, \bibinfo{year}{1952}.
\newblock \bibinfo{title}{Leaf movements in mimosa pudica l.}
\newblock \bibinfo{journal}{The New Phytologist} \bibinfo{volume}{50},
  \bibinfo{pages}{357--382}.
\bibitem[{Wijegunawardana and Amarasinghe(2021)}]{Wijegunawardana2021MEMSIVF}
\bibinfo{author}{Wijegunawardana, I.}, \bibinfo{author}{Amarasinghe, Y.W.R.},
  \bibinfo{year}{2021}.
\newblock \bibinfo{title}{The role of mems in in-vitro-fertilization}.
\newblock \bibinfo{journal}{Advances in Technology} \bibinfo{volume}{1},
  \bibinfo{pages}{235--255}.
\newblock \DOIprefix\doi{10.31357/ait.v1i1.4847}.
\bibitem[{Will et~al.(2007)Will, Tjallingii, Th{\"o}nnessen and van
  Bel}]{will2007molecular}
\bibinfo{author}{Will, T.}, \bibinfo{author}{Tjallingii, W.F.},
  \bibinfo{author}{Th{\"o}nnessen, A.}, \bibinfo{author}{van Bel, A.J.},
  \bibinfo{year}{2007}.
\newblock \bibinfo{title}{Molecular sabotage of plant defense by aphid saliva}.
\newblock \bibinfo{journal}{Proceedings of the National Academy of Sciences}
  \bibinfo{volume}{104}, \bibinfo{pages}{10536--10541}.
\bibitem[{Williams and Gilbert(1981)}]{williams1981insects}
\bibinfo{author}{Williams, K.S.}, \bibinfo{author}{Gilbert, L.E.},
  \bibinfo{year}{1981}.
\newblock \bibinfo{title}{Insects as selective agents on plant vegetative
  morphology: egg mimicry reduces egg laying by butterflies}.
\newblock \bibinfo{journal}{Science} \bibinfo{volume}{212},
  \bibinfo{pages}{467--469}.
\bibitem[{Zhang et~al.(2022)Zhang, Zhu, Gong and Zhu}]{zhang2022abiotic}
\bibinfo{author}{Zhang, H.}, \bibinfo{author}{Zhu, J.}, \bibinfo{author}{Gong,
  Z.}, \bibinfo{author}{Zhu, J.K.}, \bibinfo{year}{2022}.
\newblock \bibinfo{title}{Abiotic stress responses in plants}.
\newblock \bibinfo{journal}{Nature Reviews Genetics} \bibinfo{volume}{23},
  \bibinfo{pages}{104--119}.
\bibitem[{Zhang et~al.(2019)Zhang, Wang, Liu, Dai and Sun}]{Zhang2019}
\bibinfo{author}{Zhang, Z.}, \bibinfo{author}{Wang, X.}, \bibinfo{author}{Liu,
  J.}, \bibinfo{author}{Dai, C.}, \bibinfo{author}{Sun, Y.},
  \bibinfo{year}{2019}.
\newblock \bibinfo{title}{Robotic micromanipulation: Fundamentals and
  applications}.
\newblock \bibinfo{journal}{Annual Review of Control, Robotics, and Autonomous
  Systems} \bibinfo{volume}{2}, \bibinfo{pages}{181--203}.
\newblock
  \DOIprefix\doi{https://doi.org/10.1146/annurev-control-053018-023755}.

\end{thebibliography}

\newpage
\clearpage
\section*{Materials and Methods}\label{sec:Methods}
\subsection*{The micromanipulator}\label{sec:Methods_Robot}
All materials required for the building, assembly, and operation of the micromanipulator are collected on GitHub. These include everything needed for the hardware and software: \url{http://www.autoplant.org/AUTOPLANT/tree/main/Multi-Physics}\\
1. A complete parts list including the \textit{.stl} files for the 3D printed parts.\\
2. The PCB board design and electronic schematics to control the axis.\\
3. The electronic schematics for the electrode and laser probes.\\
4. The code to upload to the motor drivers.\\
5. The code to upload to the electrode and laser probes.\\
6. The web-based application that combines the control of the motors, probes and camera.

\subsection*{Probes}
\subsubsection*{Rod}
Inspired by earlier work by \citet{Howell2023PavementCells}, a borosilicate glass rod, with an outer diameter of 1.0 mm (BR-100-10, Sutter Instruments, USA), was placed in a micropipette puller (Model P-2000, Sutter Instruments, USA). The program values of the puller were chosen as follows: heat 400, filament 1, velocity 25, delay 160, and pull 70. The tip of each resulting probe was held over the flame of a lighter until it was dulled. This resulted in tip diameters of approximately 0.2-0.4 mm. The glass capillary was inserted into a brass drill chuck (Drill Collect Size 1.0 mm and Clamping Size 3.17 mm, Mesee, China), which, in turn, was attached to the micromanipulator.

\subsubsection*{Electrode}
The PCB from a commercially available electrical lighter (Skandi, bought in the store Kvickly, Denmark; rechargeable, 3.7 V battery) was extracted, the manual switch was exchanged with an electric relay (1 Channel Relay Module, 6.75 kV, Electrely, Amazon, Germany) and cables were exchanged to extend the reach. The relay was hooked up to an Arduino Nano Board (Arduino, Italy), which could be used to control the electrode probe from the computer. The output cables were attached to two needles (NN-2719R 27 G x 3/4", 0.4 x 20 mm, TERUMO, Japan), which were in turn placed on a 3D printed holder to mount to the manipulator. The probe tips were spaced about 0.7-1.3 mm apart, and the voltage between them was approximately 6.75 kV. The control code and 3D printed design can be found on our GitHub.

\subsubsection*{Laser}
A green LED of wavelength 520 nm (5.5-6 V green laser dot module diode, Ctricalver-DE, China) was attached to an external power source (AM-002, Input 100 V - 240 V Adjustable, 3 A max, Meiyue, China) at 5 V via an electric relay (1 Channel Relay Module, 5V, Electrely, Amazon, Germany) which in turn was connected to an Arduino Nano (Arduino, Italy) and could therefore be switched on and off via the computer. The LED was focused using a 10x microscope lens (10/0.25; 160/0.17 Achromatic Lens, Walfront, Chasoe, China). The connection between the laser probe and the micromanipulator was established through a 3D print, which can be found on our GitHub. The low-power laser probe was used with a voltage of 5 V, which resulted in a laser electrical power input of 6 mW and a maximum irradiance of 8000 W/m$^2$ for a spot size of 0.5 mm, and a maximum irradiance of 500 W/m$^2$ for a spotsize of 2 mm. The high-power laser probe was used with a voltage of 12.5 V, resulting in a laser electrical power input of 11 mW and a maximum irradiance of 3500 W/m$^2$ for a spotsize of 1 mm. For the high-power laser probe a 2.5x microscope lens (2.5x/0.075, Plan-Neoflua, Zeiss, Germany) was used for focussing instead of the 10x lens, in order to achieve a better working distance.

\subsubsection*{Needle}
Inspired by earlier work by \citet{Bae2021NeuralNetworks}, borosilicate glass with an inner and outer diameter of 0.50 mm and 1.0 mm, respectively (B100-50-10, Sutter Instruments, USA), was placed in a micropipette puller (Model P-1000, Sutter Instruments, USA). The program values of the puller were chosen as follows: heat ramp test value, pull 100, velocity 30, time 250, and pressure 500, resulting in a tip of approximately 0.02 mm in diameter. The needle tip was inserted into a 10 mM solution of 5(6)-Carboxyfluorescein diacetate (21879-25MG-F, Sigma Aldrich, USA) while applying a negative pressure at the end of the probe. Thereby, the dye was taken in through the tip. In case the tip was closed and no dye entered, then the tip was carefully broken by gently rolling it on a Lens paper (No. 1019, Linsenpapier, Assistent, Germany). The glass capillary was inserted into a brass drill chuck (Drill Collect Size 1.0 mm and Clamping Size 3.17 mm, Mesee, China) which in turn was attached to the micromanipulator.

\subsection*{Plant Material}
\textit{Arabidosis thaliana} (\textit{35S::GCaMP3}) seeds were grown in a 16h/8h day/night cycle at 22 degrees celsius for a few weeks until one or two sets of true leaves emerged.

\textit{Rosmarinus officinalis L.}, \textit{Ocimum basilicum} and \textit{Mimosa pudica} were purchased in commercial greenhouses. The plants were divided, and each stem was grown separately at room temperature without humidity and light control.

\subsection*{Calcium Measurements}
A potted \textit{Arabidosis thaliana} (\textit{35S-GCAMP}) plant was placed underneath an MZ8 Stereo Microscope (Leica, Germany) equipped with a GFP 470nm filter, a CoolLED pE-100 470nm external light source and a Leica DFC300FX camera. The plant was illuminated with the LED at 100 \% and images were taken for up to 50 min at 1 s exposure time, with a gain of 1, a gamma value of 0.6 and the highest possible frame rate, using the Leica LASX software. The plant was then exposed to the probes. A control measurement was conducted without exposure of the plant to one of the probes.

All probes were mounted on the micromanipulator and brought to the desired position using the movement system and visual inspection as a guide. The rod was brought into direct contact with a leaf every 2 min. The movement of the rod was conducted using the micromanipulator. The electrode probe was applied a few millimeters above a leaf every 20 min. The laser was turned on every 2 min for about 30 s.

Data treatment was conducted using python3. The raw images were cropped to the region of interest, and the background was subtracted by adjusting the contrast, brightness, and range of intensities until only slight autofluorescence was visible at the start of the measurement. Mean intensities were calculated for each frame, $I$, and the overall highest intensity observed, $I_{\mathrm{max}}$ was used to obtain and plot the relative intensity data over time. Time points with high intensities due to the electrode and laser probe were excluded from the graphs.

\subsection*{Microinjection}
A leaf of \textit{Rosmarinus officinalis L.} (or \textit{Ocimum basilicum}) was clipped of the full plant and fixed on a glass slide with the underside of the leaf pointing up. The needle probe filled with \textit{Fluorescin} was mounted to the micromanipulator and positioned under a microscope (DM2500, Leica, Germany). Illumination is done from above by placing an external brightfield light source (AL-M9 Amaran, Aputure, China). Fluorescent images are taken by placing a GFP 470 nm filter and illuminating from the top through an optical fiber (sola light engine, lumencor, USA). All images were taken using a colored camera (acA1920-150uc, Basler, Germany) with 150 ms exposure time, gain 0, gamma 1, fps 2. After identification of a glandular trichome, the micromanipulator is used to move the needle in position and pierce the structure. Brightfield images and fluorescent images of the field of view are taken before and after injection.

\subsection*{\textit{Mimosa pudica} Response}
A potted \textit{Mimosa pudica} was placed under direct illumination with brightflield light (AL-M9 Amaran, Aputure, China). A colored camera  (acA1920-150uc, Basler, Germany) with 150 ms exposure time, gain 0, gamma 1, fps 2 and an external lens (AF-S Nikkor 18-55mm 1:3.5-5.6G, Nikon, Japan) for imaging the plant from above. The electrode probe mounted to the micromanipulator was brought into close proximity of the plant, using visual inspection to bring it as close as possible without touching the plant and without trapping part of the plant between the two electrodes. The spark was applied and images were acquired.

The images were background corrected and the image of the field of view during the application of the electrical arc was overlayed with images after a certain time $t$ when a reaction was visible. This was achieved using a script written in python3.

\appendix
\renewcommand\thefigure{A\arabic{figure}}
\setcounter{figure}{0}
\section{Figures}
\begin{figure*}[htb]
\includegraphics[]{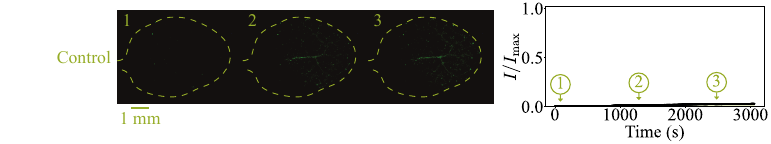}
\caption{\label{fig:CalciumControl}Calcium response without external stimulus. The fluorescence of \textit{35S:GCaMP3 Arabidopsis} was measured under illumination with blue light. We show the relative intensity $I/I_{\mathrm{max}}$ over time. A slight increase in intensity is visible, however, not very notable in the scale and compared to the experiments in Fig. \ref{fig:CalciumResponse}. Three exemplary time points were chosen to show the background-corrected images of the region of interest.}
\end{figure*}

\begin{figure*}[htb]
\includegraphics[]{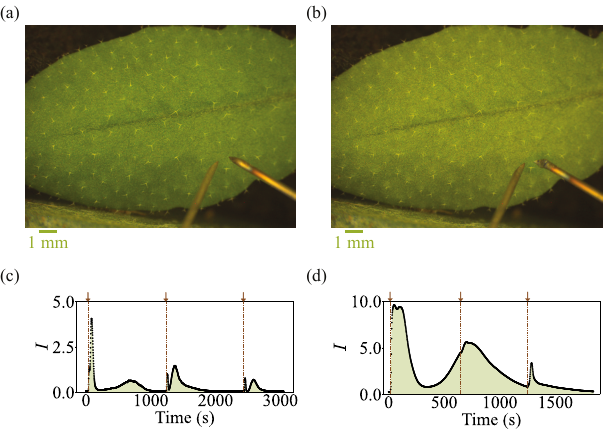}
\caption{\label{fig:CalciumBeforeAfter}Controls for calcium imaging with the electrode probe. (a-b) An Arabidopsis leaf (a) before and (b) after stimulation with the electrode probe, shows no visible tissue damages, or trichome damage around the probe. However, this does not exclude cell death on a smaller scale. (c-d) Intensity plots for shocking the same plant again 2.5 h after the initial stimulation. (c) The initial experiment with three stimuli showing a decreasing signal for each new stimulus. (d) The experiment 2.5 h after the initial experiment was started shows a recovery of the signal for the first stimulus and the same trend for the next stimuli.}
\end{figure*}

\end{document}